\documentclass[aps,epsf,preprint]{revtex4}
\usepackage{color}

\def\[{\left[}
\def\]{\right]}
\def\be{\begin{eqnarray}}
\def\ee{\end{eqnarray}}
\def\bm{\begin{pmatrix}}
\def\em{\end{pmatrix}}
\def\nn{\nonumber}
\def\({\left(}
\def\){\right)}

\def\eq#1{(\ref{#1})}
\def\a{\alpha}
\def\s{\sigma}
\def\e{\epsilon}

\def\l{\lambda}

\def\x{\times}

\def\d{\delta}
\def\labels#1{\label{#1}}
\def\bn{\begin{enumerate}}

\def\en{\end{enumerate}}
\def\b{\beta}
\def\g{\gamma}
\def\ba{\begin{array}}
\def\ea{\end{array}}
\def\bc{\begin{center}}
\def\ec{\end{center}}

\def\.{\!\cdot\!}

\def\+{\!+\!}
\def\-{\!-\!}

\def\M{M\"obius\ }
\def\Q{\Psi}
\def\pf{{\rm Pf}}

\usepackage{amssymb}
\usepackage{graphicx}

\begin{document}
\title{Off-Shell CHY Amplitudes}
\author{C.S. Lam$^a$ and and York-Peng Yao$^b$}
\affiliation{$^a$Department of Physics, McGill University\\
 Montreal, Q.C., Canada H3A 2T8\\
$^a$Department of Physics and Astronomy, University of British Columbia,  Vancouver, BC, Canada V6T 1Z1 \\
$^b$Department of Physics, The University of Michigan
Ann Arbor, MI 48109, USA\\
Emails: Lam@physics.mcgill.ca\\   yyao@umich.edu}
\title{Off-Shell CHY Amplitudes}

\begin{abstract}
The Cachazo-He-Yuan (CHY) formula for on-shell scattering amplitudes are extended off-shell. The off-shell amplitudes
 (amputated Green's functions)
are M\"obius invariant, and have the same momentum poles as the on-shell amplitudes. 
The working principles which drive the modifications to the scattering equations are mainly
M\"obius covariance and energy momentum conservation in off-shell kinematics.  The
same technique is also used to obtain off-shell massive scalars. A simple off-shell extension
of the CHY gauge formula which is M\"obius invariant is  proposed, but its true nature awaits further study.

\end{abstract}

\maketitle
\section{Introduction}
S-matrix theory was very popular in the late 1950's and early 1960's. It sought to deal more directly
with physical observables, and to avoid ultraviolet divergences by 
staying away from local
space-time interactions. Unfortunately, it never got too far because  dynamics 
could not be fully introduced without a Lorentz-invariant interaction Lagrangian density. 
This problem is now nicely circumvented by the
Cachazo-He-Yuan (CHY) scattering theory \cite{CHY1,CHY2,CHY3,CHY4,CHY5,CHY6}, where local Lorentz invariance is 
supplemented by
M\"obius invariance of the scattering amplitude in an underlying complex plane. Since its inception, there has
been many other papers discussing the properties of the scattering equations  
\cite{SE1,SE2,SE3,SE4,SE5,SE6,SE7,SE8,SE9,SE10}, 
calculations of the amplitude  \cite{CA1,CA2,CA3,DG,CA4},
its relation to string theory \cite{ST1,ST2,ST3}, the soft and collinear limits \cite{SOFT}, and
generalization to include massive and/or other particles \cite{DG,MASS1,MASS2,MASS3,MASS4}.
The CHY formula, in its original form, is a tree amplitude for massless particles. In order to implement unitarity, 
generalization to loop amplitudes \cite{LOOP3,LOOP4,LOOP5,LOOP6,LOOP1,LOOP2,LOOP7,LOOP8} is required. To facilitate such a generalization and to understand better
its connection with local quantum field theory, it is necessary to study the off-shell behavior of these scattering amplitudes.
This is what we propose to do in this paper.  In Sec.~II, we will extend the CHY on-shell scalar amplitude off-shell to 
get the amputated Green's functions.
We will also use the same technique to extend massless amplitudes to massive scalar 
amplitudes in Sec.~III, on-shell and off-shell. The on-shell version agrees with the result obtained previously by
Dolan and Goddard \cite{DG}. The same consideration also yields an off-shell extension of the CHY gauge amplitude, which is M\"obius invariant,
but the implication of such an extension requires more study as we shall discuss in Sec.~IV. Some of the illustrative details are contained
in the three appendices.

\section{Off-Shell Massless Scalar Amplitude}
Consider a set of scalar fields $\phi^{ia}$ in which the first index is
in the adjoint representation of some Lie algebra and the second index 
is in another.  If they interact tri-linearly through 

\be L_{int}={1\over 3!}f_{ijk}g_{abc} \phi^{ia}\phi^{jb}\phi^{kc}, \labels{2-1}\ee
$f$'s and $g$'s being the structure constants in the Lie algebras, then the Green's function for $n$ particle with momenta $k_i, i=1\cdots n$
at the tree level will be a function of products of propagators 
 $1\over s_{i_1i_2\cdots i_m}$ with $2\le m\le n-2$
and $s_{i_1i_2\cdots i_m}
\equiv (k_{i_1}+k_{i_2}+\cdots +k_{i_m})^2$. The coefficients will be a product $C_i$
of $n-2\ f$'s of the first Lie algebra and  another product $D_a$ of  $n-2\ g$'s of the other.  For some subsets of indices, they satisfy the Jacobi identities

\be C_i+C_j+C_k=0, \ \ \ D_a+D_b+D_c =0. \labels{2-2}\ee
Because of this and because of  $f$ and $g$ being
totally antisymmetric, only $(n-2)!$ of the $C$'s and $(n-2)!$ of the $D$'s are
independent.  We can choose an independent set, such that $C$'s are of the form 
$f_{i_1j_2j_3}\cdots f_{j_{n-2}j_{n-1}i_n}$ and $D$'s
$g_{a_1b_2b_3}\cdots g_{b_{n-2}b_{n-1}a_n}$.  The $n$-particle Green's 
function will be given as

\be \langle(\phi^{i_1a_1}(k_1) \phi^{i_2a_2}(k_2)\cdots \phi^{i_na_n}(k_n))\rangle
\approx \langle C|M|D\rangle ,  \labels{2-3}\ee
irrespective of whether $k_i$ are on-shell or not.  Here $\langle C| $ is a vector formed from the independent set just mentioned and so is
the vector $|D\rangle$.  $M$ is the $(n-2)!\times (n-2)!$ symmetric propagating  matrix given by CHY formula when all the $k_i^2=0.$  Explicit expressions for $n=4, 5$
 were given earlier by Vaman and Yao \cite{VY}.  In the next two sections, we shall explicitly solve for the modifications to the scattering functions $f_i$ in
the CHY formulas  such  that $M$ takes exactly the same form even when $k_i^2\ne 0$ for $n=4, 5$.  Generalization to any $n$ will then be given in what follows.

\subsection{Four particles}

When all the particles are on shell, the amplitudes are given by \cite{CHY1,CHY2,CHY3,CHY4,CHY5,CHY6}

\be M^{1234 \ 1ij4}=-{1\over 2\pi i}\oint {d\sigma_3\over f_3}{\sigma_{124}^2\over
\sigma_{1234}\sigma_{1ij4}},  \labels{2-4}\ee
with
\be \sigma_{i_1i_2i_3\cdots i_m}\equiv  \sigma_{i_1i_2}\sigma_{i_2 i_3}
\cdots \sigma_{i_m i_1}, \ \ \ \sigma_{ij}=\sigma_i -\sigma_j, \labels{2-5}\ee
and the scattering function $f_i$ is defined in \eq{f}. 
The integrals are to be evaluated at the pole due to $f_3=0$, with 
$\sigma_{ij}=\sigma_i-\sigma_j, \ \sigma_1=0, \ \sigma_2=1$ and $\sigma_4 \to \infty.$  We need
to evaluate the integrals for these configurations only, because from Bose 
statistics, we are required  to obtain
\be M^{1324 \  1324}=M^{1234, \ 1234}|_{ 2\leftrightarrow 3}, \ \ 
 M^{1324 \  1234}=M^{1234, \ 1324}|_{ 2\leftrightarrow 3}. \labels{2-6}\ee
We shall make changes for $f_3$ in eq.\eq{2-4} to give correct $M^{1234, 1234}$,
and $M^{1234,\ 1324}$ off-shell.  The other configurations will be given by the
substitutions of eq.\eq{2-6}.  To avoid confusion, we will use $\hat f_3$ to denote the modified $f_3$.
  The modifications we propose are

\be \hat f_3={s_{31}+x_{31}\over \sigma_{31}}+{s_{32}+x_{32}\over \sigma_{32}}, \labels{2-7}\ee
in which $x_{31}$ and $x_{32}$ are 
independent of $\sigma $'s.  This assumption of $\sigma $ independence is predicated by our preference not having to solve a high order algebraic equation  for 
the poles embedded in $f_3$ otherwise.   As we shall
see, with this assumption, Mobius covariance and energy momentum conservation
for off-shell kinematics will lead to the determination of $x_{ij}$.  However, we shall
obtain $x_{31}$ and $x_{32}$ here directly by demanding that the Green's functions from eq.(4)
should be the same as those given by the double-color field theory.
We postpone to Appendix A to show that such modifications will abide by
Mobius invariance, which allows us to set the three $\sigma$'s to the values
we gave.
The pole is at

\be \sigma_{31}=\sigma_3={s_{31}+x_{31}\over s_{31}+s_{32}+x_{31}+x_{32}}, \labels{2-8}\ee
and therefore
\be\sigma_{32}=\sigma_3-1=-{s_{32}+x_{32}\over s_{31}+s_{32}+x_{31}+x_{32}}. \labels{2-9}\ee
These give 

\be M^{1234, \ 1234}={1\over s_{31}+s_{32}+x_{31}+x_{32}}\({s_{31}+x_{31}
\over s_{32}+x_{32}} \). \labels{2-10}\ee

Using the off-shell kinematics

\be s_{31}+s_{32}+s_{12}=\sum_{i=1}^4 k_i^2, \ s_{32}=s_{14}, \labels{2-11}\ee
we obtain

\be M^{1234, \ 1234}={1\over s_{14}+x_{32}}+{1\over s_{12}-\sum _{i=1}^4 k_i^2
-(x_{31}+x_{32})}.  \labels{2-12}\ee
To coincide with the field theory result $M^{1234, \ 1234}={1\over s_{14}}+ 
{1\over s_{12}}$, we need

\be x_{31}=-\sum _{i=1}^4 k_i^2, \ \ \ x_{32}=0.   \labels{2-13}\ee  
With these, it is easy to obtain 

\be M^{1234, \ 1324}=-{1\over s_{14}}. \labels{2-14}\ee
 If we are to use these $x_{ij}$ for $\hat f_3$ and assume that $M^{1324, 1324}$ is given
by eq.(4), with $\sigma_{1234}\sigma_{1ij4}$ replaced by $\sigma_{1324}\sigma_{1324}$,
we would obtain  $M^{1324, 1324}={1\over s_{32}}+{1\over s_{31}+x_{31}}$, which is not
what field theory gives.  Instead, we should use  eq.\eq{2-6} 

\be M^{1324, \ 1324}={1\over s_{13}}+ {1\over s_{14}}, \ \ \ 
M^{1324, \ 1234}=-{1\over s_{14}}. \labels{2-15}\ee
We have not been able to find one single universal $\hat f_3$, which can produce
all the results we want for all configurations for off-shell Green's functions.
This color dependence of the off-shell scattering function is a new feature that will be further
discussed in Sec.~IIC.

\subsection{Five particles}
The amplitudes are given by \cite{CHY1,CHY2,CHY3,CHY4,CHY5,CHY6}

\be M^{12345, \ 1ijk5}=\({-1\over 2\pi i}\)^2\oint {d \sigma_2 d\sigma_4\over f_2 f_4}
{\sigma_{135}^2\over \sigma _{12345}\sigma_{1ijk5}},\labels{(2-16)}\ee
in which  $i,j,k =2,3,4$ or their permutations, and $\s_{1,3,5}$ can take on any fixed values.
The scattering functions $f_i$ are defined in \eq{f}.  
  We shall obtain the other configurations $M^{1lmn5, \ 1ijk5}$
by appropriately relabeling indices of the results from eq.\eq{(2-16)}, {\it e.g.}

\be M^{13425, \ 14235}=M^{12345,\ 13425}|_{3\to 4, 4\to 2, 2\to 3}.  \labels{(2-17)}\ee
The integrals are to be evaluated at the poles due to $f_2=0$ and $f_4=0$
simultaneously.  

In this case, let us assume that to obtain the Green's function, the modifications are

\be \hat f_2={\hat s_{21}\over \sigma_{21}}+{\hat s_{23}\over \sigma_{23}}+
    {\hat s_{24}\over \sigma_{24}}+{\hat s_{25}\over \sigma_{25}}, \labels{(2-18)}\ee
and

\be \hat f_4={\hat s_{41}\over \sigma_{41}}+{\hat s_{42}\over \sigma_{42}}+
{\hat s_{43}\over \sigma_{43}}+{\hat s_{45}\over \sigma_{45}}, \labels{(2-19)}\ee
in which $\hat s_{ij}=s_{ij}+x_{ij}$ and $x_{ij}=x_{ji}$ are assumed to be
independent of $\sigma $.

Then, for the Green's function
$M^{12345, \ 12345},$ the expected result from field theory is \cite{VY3} 

\be M^{12345, \ 12345}|_{exp}={1\over s_{15}s_{34}}+{1\over s_{15}s_{23}} +{1\over s_{12}s_{45}}+{1\over s_{23}s_{45}}+{1\over s_{12}s_{34}}.  \labels{(2-20)}\ee
From Appendix B, we find that
\be M^{12345, \ 12345}={1\over \hat s_{23}+\hat s_{34}+\hat s_{24}}({1\over \hat s_{34}}+{1\over \hat s_{23}}) +{1\over \hat s_{12}\hat s_{45}}+{1\over \hat s_{23}\hat s_{45}}+{1\over \hat s_{12}\hat s_{34}}.  \labels{(2-21)}\ee
Comparing the two equations above, we conclude that
\be s_{12}=\hat s_{12}, \ s_{45}=\hat s_{45}, \ s_{23}=\hat s_{23}, \ s_{34}=\hat s_{34},
\labels{(2-22)}\ee
which gives
\be x_{12}=x_{23}=x_{34}=x_{45}=0. \labels{(2-23)}\ee
Furthermore, we have
\be \hat s_{23}+\hat s_{34}+\hat s_{24}=s_{23}+s_{34}+s_{24}+x_{24}=s_{15}.
\labels{(2-24)}\ee
Upon using kinematics
\be s_{23}+s_{34}+s_{24}=s_{15}+k_2^2+k_3^2+k_4^2, \labels{(2-25)} \ee
we arrive at 
\be x_{24}=-(k_2^2+k_3^2+k_4^2).  \labels{(2-26)}\ee
Picking other pairs of $\hat f_i,\ \hat f_j$ to evaluate $M^{12345, \ 12345}$, we 
also obtain
\be x_{15}=0,\   x_{14}=-(k_1^2+k_4^2+k_5^2), \  x_{13}=-(k_1^2+k_2^2+k_3^2),
\labels{(2-27)}\ee
 \be x_{25}=-(k_1^2+k_2^2+k_5^2), \  x_{35}=-(k_3^2+k_4^2+k_5^2).
\labels{(2-28)}\ee

\subsection{Any number of particles}
The CHY on-shell amplitude for $n$ scalar particles is \cite{CHY1,CHY2,CHY3,CHY4,CHY5,CHY6}
\be
M^{\a,\b}=\(-{1\over 2\pi i}\)^{n-3}\oint_\Gamma\s_{rst}^2\(\prod_{i=1,i\not=r,s,t}^n{d\s_i\over f_i}\){1\over\s_\a\s_\b},\labels{m}\ee
where $\s_r,\s_s,\s_t$ are three \M constants which will be left arbitray, $\a=[\a_1\a_2\cdots\a_n]$ and 
$\b=[\b_1\b_2\cdots\b_n]$ are the two configurations of colors, with $\s_\a=\prod_{a=1}^n\s_{\a_a\a_{a+1}},\
\s_\b=\prod_{b=1}^n\s_{\b_b\b_{b+1}}$, and  $n+1\equiv 1$.
The contour $\Gamma$ encloses the $(n-3)!$ zeros of 
$f_i$ anti-clockwise, and
the on-shell scattering functions are 
\be
f_i=\sum_{j=1,j\not=i}^n{2k_i\.k_j\over\s_{ij}},\quad (1\le i\le n).\labels{f}\ee

To get an off-shell amplitude, we assume  the only change needed is to replace $f_i$ by an off-shell version given by
\be
\hat f_i&=&\sum_{j=1,j\not=i}^n{\hat s_{ij}\over\s_{ij}},\quad (1\le i\le n),\labels{hatf}\nn\\
\hat s_{ij}&=&s_{ij}+x_{ij}=2k_i\.k_j+k_i^2+k_j^2+x_{ij},\ee
where $x_{ij}=x_{ji}$ is assumed to be $\s$-independent. We also assume that
the contour $\Gamma$ is replaced by $\hat\Gamma$ to enclose $\hat f_i=0\ (i\not=r,s,t)$ anti-clockwise.
Since $x_{ii}$ do not appear, we may and will set them equal to zero. The rest of the parameters $x_{ij}=x_{ji}$ are determined
 by requiring the off-shell amplitudes to be \M invariant, and to have propagators identical to those
given by field theory.

Under a \M transformation $\s_j\to (\a\s_j+\b)/(\g\s_j+\d)$, with $\a\d-\b\g=1$, the off-shell amplitude is invariant if
$\hat f_i\to \hat f_i(\g\s_i+\d)^2$. A simple calculation shows that this is fulfilled if 
\be \sum_{j\not=i}\hat s_{ij}=0,\labels{ssum}\ee
 which,
by using momentum conservation, implies
\be \sum_{j=1}^nx_{ij}=-(n-4)k_i^2-\sum_{j=1}^nk_j^2.\labels{modular}\ee

There are many solutions to this equation, so \M invariance alone is too general to fix the off-shell amplitude, and that is where
the propagator requirement mentioned two paragraphs above comes in. First
consider the case  $\a=\b=[123\cdots n]$. Then any $(\ell\+1)\ (1\le \ell\le n-3)$ consecutive lines may form a propagator,
with an inverse factor
\be
s_{i,i+1,i+2,\cdots,i+\ell}=\sum_{r=i}^{i+\ell-1}\sum_{t=r+1}^{i+\ell}s_{rt}-(\ell-1)\sum_{r=i}^{i+\ell}k_r^2, \labels{prop}\ee
for some $i$ and some $\ell$. Here and after the line indices are understood 
to be mod $n$. For on-shell amplitudes, the inverse propagators
$\sum_{r<t}2k_r\.k_t$ can be obtained by carrying out the integration of eq.\eq{m}.
For off-shell amplitudes, the only change is to replace $f_i$ in eq.\eq{m} by $\hat f_i$, namely, by
replacing $2k_r\.k_t$ by $\hat s_{rt}$, hence the inverse propagator is 
$\sum_{r<t}\hat s_{rt}$. Equating this with eq.\eq{prop}, we get
\be \sum_{r=i}^{i+\ell-1}\sum_{t=r+1}^{i+\ell}s_{rt}-(\ell-1)\sum_{r=i}^{i+\ell}k_r^2=\sum_{r=i}^{i+\ell-1}
\sum_{t=r+1}^{i+\ell}(s_{rt}+x_{rt}),\labels{prop2}\ee
or equivalently,
\be \sum_{r=i}^{i+\ell-1}\sum_{t=r+1}^{i+\ell}x_{rt}=-(\ell-1)X_i^{i+\ell},
\quad X_i^{i+\ell}=\sum_{r=i}^{i+\ell}k_r^2,\quad (1\le \ell\le n-3).\labels{x}\ee
In particular, if $\ell=1$, then
\be x_{i,i+1}=0,\labels{m1}\ee
and
\be x_{i,i+n-1}=x_{i+n-1,i+n}=x_{i-1,i}=0.\labels{mn1}\ee
To obtain  solutions for other $x_{ij}$, subtract eq.\eq{x} from the same equation with  $\ell$ replaced by $\ell-1$  to get
\be \sum_{r=i}^{i+\ell-1}x_{r,i+\ell}=-(\ell-1)X_i^{i+\ell}+(\ell-2)X_i^{i+\ell-1},\quad (2\le \ell\le n\-3).\labels{xc1}\ee
For $\ell=2$, it gives
\be
x_{i,i+2}=-X_{i}^{i+2}=-(k_i^2+k_{i+1}^2+k_{i+2}^2),\quad(n\ge 5),\labels{m2}\ee
and
\be
x_{i,i+n-2}=x_{i+n-2,i}=x_{i+n-2,i+n}=x_{i-2,i}=X_{i+n-2}^{i+n}=-(k_{i+n-2}^2+k_{i+n-1}^2+k_{i+n}^2),\quad (n\ge 5).\nn\\  \labels{mn2}\ee
The restriction $n\ge 5$ comes about because the requirement that $2=\ell\le n-3$.
In case $n=4$, eqs.\eq{m2} and \eq{mn2} are no longer valid. They would be replaced by a relation obtained from
eqs.\eq{modular} and \eq{m1} to be
\be x_{i,i+2}=-\sum_{r=1}^4k_r^2,\quad (n=4).\ee
This agrees with the result \eq{2-13} obtained previously by direct calculation.

For $\ell\ge 3$, the solution can be obtained by subtracting \eq{xc1}
from the same equation with $(i,\ell)$ replaced by $(i\+1,\ell\-1)$ to get
\be
x_{i,i+\ell}&=&-(\ell\-1)X_i^{i+\ell}\+(\ell\-2)X_i^{i+\ell-1}\+(\ell\-2)X_{i+1}^{i+\ell}\-(\ell\-3)X_{i+1}^{i+\ell-1}\nn\\
&=&-(k_i^2+k_{i+\ell}^2),
\quad (3\le \ell\le n\-3).\labels{m3}\ee

To summarize, the solutions are
\be
x_{ij}=\left\{ \ba{ll}
          0 & \mbox{if $|j-i|=0$ or 1}\\
          -(k_i^2+k_{m}^2+k_{j}^2)& \mbox{if $|j-i|=2$}\\
          -(k_i^2+k_j^2) & \mbox{if $|j-i|\ge 3$}\\
         \ea\right\},\quad \a=\b=(123\cdots n),\labels{xsol}\ee
where $m$ in the middle equation is the line between $i$ and $j$.
These solutions are symmetric in $i$ and $j$, as they should be, and 
automatically satisfy the gauge-covariant condition eq.\eq{modular} because
\be
\sum_{j\not=i}x_{ij}=-X_{i-2}^i-X_i^{i-2}-\sum_{j=i+3}^{i+n-3}\(k_i^2+k_j^2\)=(n-4)k_i^2-\sum_{j=1}^nk_j^2.
\ee

More generally, if $\a=\b=[\a_1\a_2\a_3\cdots\a_n]$, then the inverse propagators allowed would be 
$s_{\a_i\a_{i+1}\a_{i+2}\cdots \a_{i+\ell}}$, and
the solution of $x_{ij}$ can be obtained from eq.\eq{xsol} by a substitution to get
\be
x^{\a,\a}_{\a_i\a_j}=\left\{ \ba{ll}
          0 & \mbox{if $|j-i|=0$ or 1}\\
          -(k_{\a_i}^2+k_{\a_m}^2+k_{\a_j}^2)& \mbox{if $|j-i|=2$}\\
          -(k_{\a_i}^2+k_{\a_j}^2) & \mbox{if $|j-i|\ge 3$}\\
         \ea\right\},\quad \a=\b=[\a_1\a_2\a_3\cdots \a_n].\labels{xsol1}\ee
where  the colors are now indicated in the superscripts.

So far we have considered diagonal colors, with $\b=\a$. In general, we should take $x_{ij}^{\a,\b}=(x^{\a,\a}_{ij}+x^{\b,\b}_{ij})/2$ to ensure  $M^{\a,\b}=M^{\b,\a}$. 
This gives the right answer because when $\a=\b$, it returns to eq.\eq{xsol1}.
When $\a\not=\b$,  an $(\ell\+1)$-line pole is present in $M^{\a,\b}$ only when $s_{\a_i\a_{i+1}\a_{i+2}\cdots \a_{i+\ell}}=
s_{\b_i\b_{i+1}\b_{i+2}\cdots \b_{i+\ell}}$, which demands the unordered set of momenta $\{k_{\a_i}, k_{\a_{i+1}},\cdots,k_{\a_{i+\ell}}\}$ to be identical
to the unordered set $\{k_{\b_i}, k_{\b_{i+1}},\cdots,k_{\b_{i+\ell}}\}$.
In that case the propagator requirement  eq.\eq{prop} and  eq.\eq{x}
are automatically satisfied.

It is interesting to note that the on-shell scattering functions $f_i$ do not depend on the colors, but the off-shell functions
$\hat f_i$ do.

\subsection{Off-shell amplitude and off-shell extension of on-shell amplitudes}
Take the on-shell amplitude in eq.\eq{m}. 
There is a way to extend $M^{\a,\b}$ such that three of the particles are off-shell while the rest are on-shell.
Let us call $r,s,t$ constant lines and the others $i\not=r,s,t$ variable lines. 
As long as all the variable lines are on-shell, {\it i.e.,} $k_i^2=0$ for all $i\not=r,s,t$,
then the amplitude is \M invariant no matter whether $k_r, k_s, k_t$ are on-shell or not. In this way we can define a 
M\"obius-invariant
amplitude using \eq{m} for up to three off-shell lines. We shall refer to this as the off-shell extension of the on-shell amplitude.

What we want to point out is that this off-shell extension is generally different from the off-shell amplitude 
considered above, by keeping all but at most three lines on-shell, because the off-shell extension amplitude may
not satisfy the propagator requirement. 

For example, suppose
 line $r$ is off-shell and a variable line $i$ is next to it in an amplitude with diagonal colors. The off-shell
amplitude satisfies the propagator requirement and gives rise to a propagator $1/s_{ir}$, whereas the corresponding
contribution from the off-shell extension amplitude is $1/2k_i\.k_r=1/(s_{ir}-k_r^2)$.

In the case of diagonal colors, the only time an off-shell extension amplitude coincides with an off-shell amplitude 
is when there is only one off-shell line, say $r$, shielded on either side of it by the other two on-shell constant lines $s$ and $t$.
In this way no variable line can get next to the off-shell line to produce a different propagator.

\section {Off-shell Massive Scalar Amplitude}
We want to explore whether the amplitude of a double-color scalar theory with mass $m$ 
can also be given by eq.\eq{m}, with $f_i$ replaced
by $\hat f_i$ of eq.\eq{hatf}, but with a different $x_{ij}$ than the massless case.
The general solution is given below in this section, but to make it more concrete and easier to
understand, explicit evaluations for $n=5$ and $n=6$ are given in Appendix C.

 To be \M invariant the condition eq.\eq{modular} must be satisfied. 
For $\a=\b=[123\cdots n]$, the inverse propagators are 
$s_{i,i+1,i+2,\cdots,i+\ell}-m^2$, so eq.\eq{prop2} should be replaced by
\be \sum_{r=i}^{i+\ell-1}\sum_{t=r+1}^{i+\ell}s_{rt}-(\ell-1)\sum_{r=i}^{i+\ell}k_r^2-m^2=\sum_{r=i}^{i+\ell-1}
\sum_{t=r+1}^{i+\ell}(s_{rt}+x_{rt}),\labels{prop3}\ee
and eq.\eq{x} should be replaced by
\be \sum_{r=i}^{i+\ell-1}\sum_{t=r+1}^{i+\ell}x_{rt}=-(\ell-1)\sum_{r=i}^{i+\ell}k_r^2-m^2:=-(\ell-1)X_i^{i+\ell}-m^2,\quad (1\le \ell\le n-3).\labels{x2}\ee
Setting $\ell=1$, we get
\be x_{i,i+1}=x_{i,i+n-1}=-m^2.\labels{m1m}\ee
Eq.\eq{xc1} is still valid in the massive case, because subtraction cancels the $m^2$ terms. Setting $\ell=2$, we get
\be
x_{i,i+2}=-X_{i}^{i+2}-x_{i,i+1}=-(k_i^2+k_{i+1}^2+k_{i+2}^2)+m^2,\quad(n\ge 5).\labels{m2m}\ee
Similarly, 
\be
x_{i,i+n-2}=X_{i+n-2}^{i+n}-x_{i,i+n-1}=-(k_{i+n-2}^2+k_{i+n-1}^2+k_{i+n}^2)+m^2,\quad (n\ge 5).\labels{mn2m}\ee
As in the massless case, $n=4$ must be treated separately. There, we need to use eq.\eq{modular} to get
\be
x_{i,i+2}=-x_{i,i+1}-x_{i,i-1}-\sum_{r=1}^4k_r^2=-\sum_{r=1}^4k_r^2+2m^2,\quad (n=4).\labels{n4m}\ee

For $\ell\ge 3$, the solution can be obtained by subtracting \eq{xc1}
from the same equation with $(i,\ell)$ replaced by $(i\+1,\ell\-1)$ to get
\be
x_{i,i+\ell}&=&-(\ell\-1)X_i^{i+\ell}\+(\ell\-2)X_i^{i+\ell-1}\+(\ell\-2)X_{i+1}^{i+\ell}\-(\ell\-3)X_{i+1}^{i+\ell-1}\nn\\
&=&-(k_i^2+k_{i+\ell}^2),
\quad (3\le \ell\le n\-3).\labels{m3m}\ee
The final solution is therefore
\be
x_{ij}=\left\{ \ba{ll}
          0 & \mbox{if $|j-i|=0$}\\
		-m^2 & \mbox{if $|j-i|=1$}\\
          -(k_i^2+k_{m}^2+k_{j}^2)+m^2& \mbox{if $|j-i|=2$}\\
          -(k_i^2+k_j^2) & \mbox{if $|j-i|\ge 3$}\\
         \ea\right\},\quad \a=\b=(123\cdots n).\labels{xsolm}\ee
It can easily be checked that the \M-invariant condition eq.\eq{modular} is also automatically satisfied. For general colors
$\a$ and $\b$, the solution can again be obtained from eq.\eq{xsolm} by a substitution as in the massless case.

\section{An Off-shell extension of the gauge Amplitude}
Similar to \eq{m}, the on-shell color-stripped $n$-gluon scattering amplitude is given by the CHY formula \cite{CHY1,CHY2,CHY3,CHY4,CHY5,CHY6} to be
\be
M^{\a}=\(-{1\over 2\pi i}\)^{n-3}\oint_\Gamma\s_{rst}^2\(\prod_{a=1,a\not=r,s,t}^n{d\s_a\over f_a}\){\pf'\Q\over\s_\a},\labels{mg}\ee
with the reduced Pfaffian $\pf'\Q$ replacing the factor $1/\s_\b$ in the scalar theory.
The reduced Pfaffian is related to the Pfaffian of $\Q^{ij}_{ij}$ by
\be \pf\Q'=2{(-1)^{i+j}\over\s_{ij}}\pf\(\Q^{ij}_{ij}\),\ee
the matrix $\Q^{ij}_{ij}$ is obtained from the matrix $\Q$ by deleting its $i$th column and row and its $j$th
column and row, and the antisymmetric matrix $\Q$ is made up of three $n\x n$ matrices $A, B, C$,
\be 
\Q=\bm{A&-C^T\cr C&B\cr}\em.\ee 
The non-diagonal elements of these three sub-matrices are
\be A_{ab}={2k_a\.k_b\over\s_{ab}},\quad B_{ab}={\e_a\.\e_b\over\s_{ab}},\quad C_{ab}={\e_a\.k_b\over\s_{ab}},\quad 1\le a
\not=b\le n,\labels{AAA}\ee
where $\e_a$ is the polarization of the $a$th gluon, satisfying $\e_a\.k_a=0$. 
The diagonal elements of $A$ and $B$ are zero, and that of $C$ is given by
\be C_{aa}=-\sum_{b=1}^nC_{ab},\labels{Csum}\ee
so that the column and row sums of $C$ is zero. A similar property is true for $A$ if the scattering equations $f_a=0$
are obeyed, which is the case because the integration contour $\Gamma$ encloses these zeros anti-clockwise. 

Under a \M transformation, $\s_b\to (\a\s_b+\b)/(\g\s_b+\d)$, with $\a\d-\b\g=1$, 
the CHY amplitude eq.\eq{mg} is \M invariant if the momenta are massless and conserved.
The gluon amplitude is gauge invariant and independent of the choice of $\l$ and $\nu$ if the row sum and
column sum of the sub-matrices of $A$ and $C$ are zero, as mentioned in \eq{Csum} and the paragraph below it.

An off-shell extension of \eq{mg} can be obtained if we replace $f_i$ by $\hat f_i$ obtained in the previous sections,
the contour $\Gamma$ by $\hat\Gamma$ enclosing $\hat f_i=0$,
and the elements of $A$ in \eq{AAA} by 
\be A_{ij}=\sum_{j\not=i}{\hat s_{ij}\over\s_{ij}}.\ee
The row and column sums of $A$ are still zero because $\hat f_i=0$ and because \eq{ssum} is satisfied. This off-shell extension
is M\"obius invariant and is independent of the choice of $\l,\nu$ as before, because all the conditions necessary to prove
these properties for the on-shell amplitude have been preserved with the change.

These changes are the simplest extensions of the on-shell scattering formula to off-shell, but  whether it is the amputated Green's function of an Yang-Mills field theory is not immediately clear. The reason is, off-shell Yang-Mills
theory is gauge dependent, and in our extension gauges do not enter. It is possible that this extension determines a particular `CHY gauge',
or that the true off-shell extension is much more complicated than what is discussed in order to reflect the the freedom of gauge choice. It is even
possible that field-theoretical off-shell expression is not M\"obius invariant. Further study is required to know what is the truth.

\section{Conclusion}
The CHY scattering formulas for massless scalar particles are extended off-shell
by changing $2k_i\.k_j$ in the scattering function$f_i$ to $(k_i+k_j)^2+x_{ij}$,
where $x_{ij}=x_{ji}$ is independent of $\s$.  It can be determined uniquely from the requirement that
off-shell amplitudes are M\"obius invariant and have exactly the same invariant-momentum poles
as the on-shell amplitudes. The same requirements also allow us to extend the formula to
massive scalar and vector amplitudes, on-shell and off-shell. 
A simple off-shell extension of the CHY gauge amplitude is also proposed, with many nice properties including \M invariance
and the independence of $\l$ and $\nu$, but the true nature of this extension formula requires further study.

\appendix
\section{\M invariance of the $n=4$ and $n=5$ amplitudes}
One motivating and intriguing feature of the CHY formulas is that the
on-shell amplitudes for scalar, gauge,  and gravitational interactions are 
all invariant under M\"obius transformations

\be \sigma_i={\alpha \sigma_i'+\beta \over \gamma \sigma_i'+\delta}, \ \ \ \ 
\alpha \delta -\beta \gamma=1. \labels{ (4-1)}\ee 
In our extending the CHY formula to off-shell, it is very natural to ask
if such invariance still holds.  In fact, this is required of us, because if it
were not so, then we would not have the  freedom to fix  three of the $\sigma's$ to the values we used in Sections IIA and IIB.  Gladly, the answer is in the affirmative,  although with some restrictions (eq.\eq{2-6}, eq.\eq{(2-17)}).  Let us begin with the case of four double-color
scalars.  The invariance of the off-shell Green's functions is intimately
tied up with the transformation property of the scattering equations.  Let 
us generalize these slightly before we fix three of the $\sigma $'s

\be  \hat f_1(\sigma )&=&{s_{12}\over \sigma_{12}}+{s_{13}+x_{13}\over \sigma_{13}}
+{s_{14}\over \sigma_{14}} ,\nn\\ 
\hat f_2(\sigma )&=&{s_{21}\over \sigma_{21}}+{s_{23}\over \sigma_{23}}
+{s_{24}+x_{31}\over \sigma_{24}} ,\nn\\
\hat f_3(\sigma )&=&{s_{31}+x_{31}\over \sigma_{31}}+{s_{32}\over \sigma_{32}}
+{s_{34}\over \sigma_{34}} ,\nn\\
 \hat f_4(\sigma )&=&{s_{41}\over \sigma_{41}}+{s_{42}+x_{31}\over \sigma_{42}}
+{s_{43}\over \sigma_{43}}. 
\labels{(4-2)} \ee
One can verify that they satisfy 
\be \sum_{i=1}^4 \hat f_i=\sum _{i=1}^4 \sigma _i \hat f_i=\sum_{i=1}^4 \sigma_i^2\hat f_i=0,
\labels{(4-3)}\ee
as a result of momentum conservation, and
\be \sum_{j\ne i} s_{ij}=\sum _{j=1}^4 k_j^2=-x_{31}=-x_{13}, \labels{(4-4)}\ee
which in turn means that only one of the $\hat f_i$ is independent.  Let 
us take this to be $\hat f_3$.  Then, we find that under M\"obius transformation

\be \hat f_3(\sigma)=\gamma (\gamma \sigma_3'+\delta )[-(s_{31}+x_{13})-s_{32}-s_{34}]
+(\gamma \sigma_3'+\delta )^2\hat f_3(\sigma ').  \labels{(4-5)}\ee
The first term vanishes as indicated by eq.\eq{(4-4)}, which makes $\hat f_3$ M\"obius covariant.   Now, we also
set $\sigma_1=0, \ \sigma_2=1$ and $\sigma _4 \to \infty$.
Using 

\be d\sigma_3={d\sigma_3'\over (\gamma \sigma_3'+\delta )^2},  \labels{(4-6)}\ee
and
\be {\sigma_{124}\over \sigma_{1234}\sigma_{1ij4}}=(\gamma \sigma_3'+\delta)^4
 {\sigma'_{124}\over \sigma'_{1234}\sigma'_{1ij4}}, \ \ \ \ i,j=2,3 , \labels{(4-7)}\ee
 we have 
 
 \be \oint {d\sigma_3\over \hat f_3(\sigma)}{\sigma_{124}\over \sigma_{1234}\sigma_{1ij4}}=
\oint {d\sigma'_3\over \hat f_3(\sigma')}{\sigma'_{124}\over \sigma'_{1234}\sigma'_{1ij4}}.
\labels{(4-8)}\ee
The caveat here is that in order to have poles at the correct place, we have
committed to the form of $\hat f_3$ as determined. 

For the double-color five particle amplitudes, 
let us generalize the functions slightly
to
\be \hat f_1&=&{s_{12}\over \sigma_{12}}+{s_{13}+x_{13}\over \sigma_{13}}+{s_{14}+x_{14}\over \sigma_{14}}+{s_{15}\over \sigma_{15}},\nn\\
\hat f_2&=&{s_{21}\over \sigma_{21}}+{s_{23}\over \sigma_{23}}+{s_{24}+x_{24}\over \sigma_{24}}+{s_{25}+x_{25}\over \sigma_{25}}, \nn\\
\hat f_3&=&{s_{31}+x_{31}\over \sigma_{31}}+{s_{32}\over \sigma_{32}}+{s_{34}\over \sigma_{34}}+{s_{35}+x_{35}\over \sigma_{35}},\nn\\
\hat f_4&=&{s_{41}+x_{41}\over \sigma_{41}}+{s_{42}+x_{42}\over \sigma_{42}}+{s_{43}\over \sigma_{43}}+{s_{45}\over \sigma_{45}}, \nn\\
\hat f_5&=&{s_{51}\over \sigma_{51}}+{s_{52}+x_{52}\over \sigma_{52}}+{s_{53}+x_{53}\over \sigma_{53}}+{s_{54}\over \sigma_{54}}.\labels{(4-9)}\ee

It is easy to check that 

\be\sum_{i=1}^5\hat f_i=\sum _{i=1}^5 \sigma _i \hat f_i=\sum_{i=1}^5 \sigma_i^2 \hat f_i=0,
\labels{(4-10)}\ee
because of momentum conservation, similar to eq.(4-4), 
when we take

\be x_{31}=&-(k_1^2+k_2^2 + k_3^2),  \ \ x_{41}=-(k_1^2+k_4^2+k_5^2),\ \ x_{42}=-(k_2^2+k_3^2+k_4^2), \cr &
\ \ x_{52}=-(k_1^2+k_2^2+k_5^2),\ \ x_{53}=-(k_3^2+k_4^2+k_5^2), \labels{(4-11)}\ee
which implies that only two of the $\hat f_i's$ are independent.   We choose  
them to be $\hat f_3$ and $\hat f_4$, which are M\"obius covariant, in the sense that

\be \hat f_3(\sigma )=(\gamma \sigma_3'+\delta )^2\hat f_3(\sigma'), \ \ \ \ 
\hat f_4(\sigma )=(\gamma \sigma_4'+\delta )^2\hat f_4(\sigma'), \labels{(4-12)}\ee
and we are led to

\be \oint {d\sigma_3 d\sigma_4\over \hat f_3(\sigma)\hat f_4(\sigma)}
{\sigma_{125}^2\over \sigma_{12345}\sigma_{1ijk5}}= 
\oint {d\sigma'_3 d\sigma'_4\over \hat f_3(\sigma')\hat f_4(\sigma')}
{(\sigma'_{125})^2\over \sigma'_{12345}\sigma'_{1ijk5}}. \labels{(4-13)}\ee

\section{A diagonal element of the $n=5$ amplitude}
 In this note one of $n=5$ scalar amplitudes is calculated. The others 
 can be done in a similar fashion.  For our purpose here, 
 let us consider the most complicated case with diagonal colors, say with 1,3,5 as the constant lines.
\be I_5&=&\({-1\over 2\pi i}\)^2\oint_\Gamma {\s_{135}^2d\sigma_2 d\sigma _4\over  f_2  f_4 \sigma_{12345}\s_{12345}},\labels{I5}\\
 f_2&=&{s_{21}\over\s_{21}}+{s_{23}\over\s_{23}}+{s_{24}\over\s_{24}}+{s_{25}\over\s_{25}},\nn\\
 f_4&=&{s_{41}\over\s_{41}}+{s_{42}\over\s_{42}}+{s_{43}\over\s_{43}}+{s_{45}\over\s_{45}}. \labels{I52}\ee
 Poles are from $\{21\}, \{23\}, \{234\}, \{43\}, \{45\}$. We denote contributions from 
 $((\{21\}, \{43\}), (\{21\}, \{45\}))$,  $(\{23\}, \{45\})$, and $\{234\}$
 by $I_{5a}, I_{5b}, I_{5c}$ respectively.  
 
 For the first case, 
 \be
 I_{5a}&=&-{1\over 2\pi i}\oint_{\Gamma_4}{\s_{135}^2d\s_4\over f_{2a}f_{4a}\s_{1345}^2},\nn\\
 f_{2a}&=&s_{21},\nn\\
 f_{4a}&=&{s_{41}+s_{42}\over\s_{41}}+{s_{43}\over\s_{43}}+{s_{45}\over\s_{45}}.
 \ee
 In the subsequent $\s_4$ integration, both the $\s_{43}=0$ and the $\s_{45}=0$ poles contribute to give two terms.
 The final result is 
 \be I_{5a}={1\over s_{21}}\({1\over s_{43}}+{1\over s_{45}}\).\ee
 
 Similarly,  for the second case
 \be I_{5b}={1\over s_{23}}{1\over s_{45}}.\ee
 
 As for $I_{5c}$, there are two regions of contribution.   Carry out the change of variables
  $\s_{23}=s\s'_{23}$ and $\s_{34}=s\s'_{34}$, from $\s_2$ and $\s_4$ to $s$ and some linear combination of
  $\s'_{23}$ and $\s'_{34}$. In the vicinity of $s=0$, the factor $\s_{12345}^2$ becomes 
  $s^4\s_{135}^2(\s'_{23}\s'_{34})^2$, which shows two zeros, the first at $\s'_{43}=0$
  and the second at $\s'_{23}=0$. In the region around the first zero, we fix $\s'_{23}=1$, so the integration
  measure  becomes 
 $d\s_2 d\s_4=sdsd\s'_4$. After the $s$ integration, we end up with
 \be I_{5c}&=&-{1\over 2\pi i}\oint_{\Gamma_4}{\s_{135}^2d\s'_4\over f_{2c}f_{4c}\s_{135}^2(\s'_{23}\s'_{34})^2},\nn\\
 f_{2c}&=&{s_{23}\over\s'_{23}}+{s_{24}\over\s'_{24}},\nn\\
 f_{4c}&=&{s_{42}\over\s'_{42}}+{s_{43}\over\s'_{43}}. \ee
 The contour $\Gamma_4$ forces $f_{4c}=0$. With $\s'_{23}=1$, this yields $\s'_{43}=s_{43}/(s_{42}+s_{43})$,
 and hence $f_{2c}=s_{23}+s_{34}+s_{24}=s_{15}$. Now reverse and distort the $\s_4'$ contour to
 surround the pole at $\s'_{43}=0$. In this way we get
 \be I^{1st}_{5c}={1\over s_{15}s_{43}}.\ee
 Contribution from the second region is obtained by $2\leftrightarrow 4$
 \be I^{2nd}_{5c}={1\over s_{15}s_{23}}.\ee
Thus $I_5=I_{5a}+I_{5b}+I_{5c}$ consists of 5 terms, corresponding to five Feynman diagrams.

\section{Off-shell Scattering Equations for $n=5, 6$ Massive Scalars}
In the following, we are going to give two examples to illustrate how to 
construct the scattering equations which will give the same amplitudes
as from field theory.  There, the massive scalar propagators 
are ${1\over s_{i_1 i_2\cdots i_p -m^2}}, \ 2 \le p  \le n-2$, instead
of ${1\over s_{i_1 i_2\cdots i_p}}$ as in the massless case.  Hence,
the obvious rule is first to replace $s_{i_1 i_2\cdots i_p}$  with  $s_{i_1 i_2\cdots i_p}
-m^2$ in the scattering equations of Section II.  Consider the case $n=5$, where
now eq.\eq{(4-2)} becomes

\be \hat f_1&=&{s_{12}-m^2\over \sigma_{12}}+{s_{13}-m^2+x'_{13}\over \sigma_{13}}
+{s_{14}-m^2+x'_{14}\over \sigma_{14}}+{s_{15}-m^2\over \sigma_{15}},\nn\\
\hat f_2&=&{s_{21}-m^2\over \sigma_{21}}+{s_{23}-m^2\over \sigma_{23}}
+{s_{24}-m^2+x'_{24}\over \sigma_{24}}+{s_{25}-m^2+x'_{25}\over \sigma_{25}},\nn\\
\hat f_3&=&{s_{31}-m^2+x'_{31}\over \sigma_{31}}+{s_{32}-m^2\over \sigma_{32}}
+{s_{34}-m^2\over \sigma_{34}}+{s_{35}-m^2+x'_{35}\over \sigma_{35}},\nn\\
\hat f_4&=&{s_{41}-m^2+x'_{41}\over \sigma_{41}}+{s_{42}-m^2+x'_{42}\over \sigma_{42}}
+{s_{43}-m^2\over \sigma_{43}}+{s_{45}-m^2\over \sigma_{45}},\nn\\ 
\hat f_5&=&{s_{51}-m^2\over \sigma_{51}}+{s_{52}-m^2+x'_{52}\over \sigma_{52}}
+{s_{53}-m^2+x'_{53}\over \sigma_{53}}+{s_{54}-m^2\over \sigma_{54}}, \labels{(B-1)}\ee
where we have written in the notation of Section III $x_{ij}=x'_{ij}-m^2$.  
Modular covariance of $\hat f_1=0$ gives the condition

\be(s_{12}-m^2)+(s_{13}-m^2+x'_{13})+(s_{14}-m^2+x'_{14})+(s_{15}-m^2)=0.
\labels{(B-2)}\ee
Using momentum conservation, we have
\be s_{12}+s_{13}+s_{14}+s_{15}=\sum_{i=1}^5 k_i^2+k_1^2, \labels{(B-3)}\ee
which results in
\be x'_{13}+x'_{14}=-({\sum_{i=1}^5k_i^2+k_1^2})+4m^2.  \labels{(B-4)}\ee
Similar consideration gives
\be x'_{24}+x'_{25}&=&-({\sum_{i=1}^5k_i^2+k_2^2})+4m^2, \nn\\
x'_{13}+x'_{35}&=&-({\sum_{i=1}^5k_i^2+k_3^2})+4m^2, \nn\\
x'_{14}+x'_{24}&=&-({\sum_{i=1}^5k_i^2+k_4^2})+4m^2, \nn\\ 
x'_{25}+x'_{35}&=&-({\sum_{i=1}^5k_i^2+k_5^2})+4m^2,\labels{(B-5)}\ee
which lead to the solution

\be x'_{13}&=&-(k_1^2+k_2^2+k_3^2)+2m^2, \nn\\
x'_{14}&=&-(k_1^2+k_4^2+k_5^2)+2m^2, \nn\\
x'_{24}&=&-(k_2^2+k_3^2+k_4^2)+2m^2, \nn\\
x'_{25}&=&-(k_1^2+k_2^2+k_5^2)+2m^2,\nn\\
x'_{35}&=&-(k_3^2+k_4^2+k_5^2)+2m^2.\labels{(B-6)}\ee
These shifts will give us the amplitudes $M^{12345, \ 1ijk5}$, where 
$i, j, k$ are any permutations of $2,3,4$.  Particularly, we have

\be M^{12345 \ 12345}&=&{1\over (s_{15}-m^2)(s_{34}-m^2)}+{1\over (s_{15}-m^2)(s_{23}-m^2)}+{1\over (s_{12}-m^2)(s_{45}-m^2)}\nn\\ 
&+&{1\over (s_{23}-m^2)(s_{45}-m^2)}
+{1\over (s_{12}-m^2)(s_{34}-m^2)}.\labels{(B-7)}\ee
Let us take note that in the scattering equations eq.\eq{(B-1)}, the set of invariants which do 
not require $x'_{ij}$ shifts other than $-m^2$ are $s_{12}, \ s_{23}, \ s_{34}, \ s_{45}, \ s_{51}$
and they all appear as propagators in $M^{12345 \ 12345}$, whereas the complement 
set consisting of $s_{13},\ s_{14}, \ s_{24}, \ s_{25} \ s_{35},$  require
non-trivial shifts.  This will prove to be true for all $n$.  With this in mind, we now look at 
$n=6$.  The invariants which appear as propagators in $M^{123456, 1ijklm6}$,
where $i, j, k, l$ are permutations of $2, 3, 4, 5$, are $s_{12}, s_{23}, s_{34},
s_{45}, s_{56}, s_{61}, s_{123}=s_{456}, s_{234}=s_{561}, s_{345}=s_{612}.$
They come with consecutive indices. 
The complement to this set is $s_{13}, s_{14}, s_{15}, s_{24}, s_{25}, s_{26}, s_{35}, s_{36}, s_{46}.$  Thus the modified scattering equations are

\be \hat f_1&=&{s_{12}-m^2\over \sigma_{12}}+{s_{13}-m^2+x'_{13}\over \sigma_{13}}
+{s_{14}-m^2+x'_{14}\over \sigma_{14}}+{s_{15}-m^2+x'_{15}\over \sigma_{15}} 
+{s_{16}-m^2\over \sigma_{16}},\nn\\
\hat f_2&=&{s_{21}-m^2\over \sigma_{21}}+{s_{23}-m^2\over \sigma_{23}}
+{s_{24}-m^2+x'_{24}\over \sigma_{24}}+{s_{25}-m^2+x'_{25}\over \sigma_{25}}
+{s_{26}-m^2+x'_{26}\over \sigma_{26}},\nn\\
\hat f_3&=&{s_{31}-m^2+x'_{31}\over \sigma_{31}}+{s_{32}-m^2\over \sigma_{32}}
+{s_{34}-m^2\over \sigma_{34}}+{s_{35}-m^2+x'_{35}\over \sigma_{35}}
+{s_{36}-m^2+x'_{36}\over \sigma_{36}},\nn\\ 
\hat f_4&=&{s_{41}-m^2+x'_{41}\over \sigma_{41}}+{s_{42}-m^2+x'_{42}\over \sigma_{42}}
+{s_{43}-m^2\over \sigma_{43}}+{s_{45}-m^2\over \sigma_{45}}
+{s_{46}-m^2+x'_{46}\over \sigma_{46}},\nn\\
\hat f_5&=&{s_{51}-m^2+x'_{51}\over \sigma_{51}}+{s_{52}-m^2+x'_{52}\over \sigma_{52}}
+{s_{53}-m^2+x'_{53}\over \sigma_{53}}+{s_{54}-m^2\over \sigma_{54}}
+{s_{56}-m^2\over \sigma_{56}},\nn\\
\hat f_6&=&{s_{61}-m^2\over \sigma_{61}}+{s_{62}-m^2+x'_{62}\over \sigma_{62}}
+{s_{63}-m^2+x'_{63}\over \sigma_{63}}+{s_{64}-m^2+x'_{64}\over \sigma_{64}}
+{s_{65}-m^2\over \sigma_{65}}.\labels{(B-8)}\ee
The requirement of M\"obius covariance leads to

\be x'_{13}+x'_{14}+x'_{15}&=&-(\sum_{j=1}^6k_i^2+2k_1^2)+5m^2,\nn\\
x'_{24}+x'_{25}+x'_{26}&=&-(\sum_{j=1}^6k_i^2+2k_2^2)+5m^2,\nn\\
x'_{13}+x'_{35}+x'_{36}&=&-(\sum_{j=1}^6k_i^2+2k_3^2)+5m^2,\nn\\
x'_{14}+x'_{24}+x'_{46}&=&-(\sum_{j=1}^6k_i^2+2k_4^2)+5m^2,\nn\\
x'_{15}+x'_{25}+x'_{35}&=&-(\sum_{j=1}^6k_i^2+2k_5^2)+5m^2,\nn\\
x'_{26}+x'_{36}+x'_{46}&=&-(\sum_{j=1}^6k_i^2+2k_6^2)+5m^2.\labels{(B-9)}\ee
Now consider a propagator which involves three momenta, such as 
${1\over s_{123}-m^2}$.  We write

\be s_{123}=s_{12}+s_{13}+s_{23}-(k_1^2+k_2^2+k_3^2),\labels{(B-10)}\ee
and acknowledge that the induced dependence on the kinematical invariants 
in the scattering amplitudes due to CHY
integrations must be in the form of those combinations $s_{ij}-m^2+x'_{ij}$ which appear in $\hat f_i$ of eq.\eq{(B-8)}.  Therefore

\be s_{123}-m^2=(s_{12}-m^2)+(s_{13}-m^2+x'_{13})+(s_{23}-m^2). \labels{(B-11)}\ee
These two equations immediately above give us a consistency condition

\be x'_{13}=-(k_1^2+k_2^2+k_3^2)+2m^2.\labels{(B-12)}\ee
Similar considerations lead to

\be x'_{24}&=&-(k_2^2+k_3^2+k_4^2)+2m^2, \ \ \ x'_{35}=-(k_3^2+k_4^2+k_5^2)+2m^2,\nn\\
x'_{46}&=&-(k_4^2+k_5^2+k_6^2)+2m^2,\ \ \ x'_{15}=-(k_1^2+k_5^2+k_6^2)+2m^2, \nn\\
   x'_{26}&=&-(k_1^2+k_2^2+k_6^2)+2m^2.\labels{(B-13)}\ee
Note that we have twelve equations in eqs.\eq{(B-9)},\eq{(B-12)},\eq{(B-13)} for nine $x'_{ij}$.
In other words, there are three consistency checks.  The rest of the solution are

\be x'_{14}=-(k_1^2+k_4^2)+m^2,\ \ x'_{25}=-(k_2^2+k_5^2)+m^2, \ \ 
x'_{36}=-(k_3^2+k_6^2)+m^2.\labels{(B-14)}\ee

When all the particles are on-shell ($k_i^2=m^2, i=1,2,\cdots n$), we find 
that all the non-trivial $x'_{ij}=-m^2$ in eqs.\eq{(B-6)}, \eq{(B-12)}, \eq{(B-13)}, \eq{(B-14)}.  This result was
obtained earlier by Dolan and Goddard \cite{DG} for $n=4, 5$.  We must reiterate that 
the scattering equations with the general values given above for $x'_{ij}$ 
should be used only for $M^{12\cdots,n \ 1 i_2 i_3 \cdots i_{n-1} n}$, where
$i_2i_3\cdots i_{n-1}$ are permutations of $2,3,\cdots ,n-1$  The other 
color amplitudes should be obtained from these by appropriate relabelling
of indices, or by following the rule we gave after eq.(49).

\end{document}